\documentclass[twocolumn,showpacs,preprintnumbers,amsmath,aps]{revtex4}
\usepackage{graphicx}
\usepackage{dcolumn}
\usepackage{bm}
\begin{document}
\def\eq#1{\ref{#1}}
\def\fig#1{\hspace{1mm}\ref{#1}}
\def\tab#1{\hspace{1mm}\ref{#1}}
\title{Strong-coupling description of the high-temperature superconductivity in the molecular hydrogen}
\author{R. Szcz{\c{e}}{\`s}niak, M.W. Jarosik}
\affiliation{Institute of Physics, Cz{\c{e}}stochowa University of Technology, Al. Armii Krajowej 19, 42-200 Cz{\c{e}}stochowa, Poland}
\email{jarosikmw@wip.pcz.pl}
\date{\today} 
\begin{abstract}
The detailed study of the selected thermodynamic properties of the superconducting phase in the molecular hydrogen under the pressure at $428$ GPa has been presented. For the increasing value of the Coulomb pseudopotential, $\mu^{*}\in\left<0.08,0.15\right>$, the following results have been obtained: (i) the critical temperature decreases from $179$ K to $141$ K, (ii) the ratio $R_{1}\equiv\Delta\left(0\right)/k_{B}T_{C}$ differs noticeably from
the BCS value: $R_{1}\in\left<4.71,3.60\right>$; (iii) the electron effective mass is large and grows slightly together with the temperature 
($\left[m^{*}_{e}/m_{e}\right]_{{\rm max}} = 2.2$ for $T = T_{C}$). 
\end{abstract}
\maketitle
%
\section{Introduction}

At low temperature, hydrogen exhibits the nontrivial structural behavior under the pressure ($p$) \cite{Narayana}, \cite{Loubeyre}. Below $110$ GPa the hexagonal-closed-packed lattice with freely rotating molecules is stable (phase I). At higher pressures, up to $150$ GPa, the broken symmetry phase has been observed (phase II). In the pressure range of $150$-$350$ GPa the so called phase III exists. The experimental measurements have proved, that the all listed phases do not demonstrate the metallic properties. In the pressure range from $400$ to $500$ GPa, the theoretical studies predict the existence of the molecular metallic phase (the {\it Cmca} crystal structure) \cite{Johnson}, \cite{Stadele}, \cite{Pickard}. Above $500$ GPa, the molecular metallic phase transforms to the Cs-IV monatomic phase \cite{Johnson}, \cite{Stadele}, \cite{Pickard}, \cite{Natoli}. This phase is stable at least up to $802$ GPa \cite{Yan}. For the extremely high value of the pressure ($2000$ GPa) Maksimov and Savrasov have proposed the simple {\it fcc} structure \cite{Maksimov}.

The molecular and monatomic metallic form of the hydrogen can be the superconductor with the high critical temperature ($T_{C}$) \cite{Ashcroft}.
In particular, the calculated values of the critical temperature have been presented in Tab.\tab{t1}. We notice, that $T_{C}$ usually has been obtained by using the McMillan formula \cite{McMillan}, which represents the weak coupling limit of the more elaborate Eliashberg approach \cite{Eliashberg}. However, in the case of the metallic hydrogen the electron-phonon interaction is strong, hence the McMillan expression is inappropriate. 

For this reason, we have calculated the critical temperature with the help of the Eliashberg equations. We have considered the case $p=428$ GPa. Additionally, we have studied precisely the properties of the order parameter and the electron effective mass. 

In the paper we have taken into consideration the Eliashberg set in the mixed representation \cite{Marsiglio}. This approach allows one to obtain the stable solutions on the real axis, since the analysis does not involve any principal-part integrals with singular integrands. 

\begin{table}
\caption{\label{t1} The critical temperature for the selected values of the pressure; $\mu^{*}$ denotes the Coulomb pseudopotential.}
\begin{ruledtabular}
\begin{tabular}{ccccc}
$p$ (GPa)& $T_{C}$ (K)& $\mu^{*}$& Structure & Ref.\\
\hline
347  & 107\footnote{The McMillan formula \cite{McMillan}.}                                    & 0.1          & {\it Cmca} &\cite{Zhang}\\
347  & (120, 90)\footnote{The exact solution of the Eliashberg equations \cite{Eliashberg}.}  & (0.08, 0.15) & {\it Cmca} &\cite{Szczesniak1}\\
388  & 130$^{a}$                                                                              & 0.1          & {\it Cmca} &\cite{Zhang}\\
400  & 130-230$^{a}$                                                                          & 0.1          & {\it sh, dsh, 9R}
\footnote{Probably unstable.} &\cite{Barbee}\\
414  & 84\footnote{The three-band model.}                                                     & -            & {\it Cmca} &\cite{Cudazzo}\\
428  & 162$^{a}$                                                                              & 0.1          & {\it Cmca} &\cite{Zhang}\\
450  & 242$^{d}$                                                                              & -            & {\it Cmca} &\cite{Cudazzo}\\
480  & 284 (266)$^{a}$                                                                        & 0.1 (0.13)   & Cs-IV      &\cite{Yan}\\
539  & 291 (272)$^{a}$                                                                        & 0.1 (0.13)   & Cs-IV      &\cite{Yan}\\
608  & 291 (271)$^{a}$                                                                        & 0.1 (0.13)   & Cs-IV      &\cite{Yan}\\
802  & 282 (260)$^{a}$                                                                        & 0.1 (0.13)   & Cs-IV      &\cite{Yan}\\
2000 & $\sim 600$                                                                             & -            & {\it fcc}  &\cite{Maksimov}\\
2000 & (631, 413)$^{b}$                                                                       & (0.1, 0.5)   & {\it fcc}  &\cite{Szczesniak3}\\
\end{tabular}
\end{ruledtabular}
\end{table}

\section{The Eliashberg equations}

The Eliashberg equations in the mixed representation have been written in the following form \cite{Eliashberg}:
\begin{eqnarray}
\label{r1}
\phi\left(\omega\right)&=&
\frac{\pi}{\beta}\sum_{m=-M}^{M}\frac{\left[\lambda\left(\omega-i\omega_{m}\right)-\mu^{*}\left(\omega_{m}\right)\right]}
{\sqrt{\omega_m^2Z^{2}_{m}+\phi^{2}_{m}}}\phi_{m}\\ \nonumber
                              &+& i\pi\int_{0}^{+\infty}d\omega^{'}\alpha^{2}F\left(\omega^{'}\right)
                                  [\left[N\left(\omega^{'}\right)+f\left(\omega^{'}-\omega\right)\right]\\ \nonumber
                              &\times&K\left(\omega,-\omega^{'}\right)\phi\left(\omega-\omega^{'}\right)]\\ \nonumber
                              &+& i\pi\int_{0}^{+\infty}d\omega^{'}\alpha^{2}F\left(\omega^{'}\right)
                                  [\left[N\left(\omega^{'}\right)+f\left(\omega^{'}+\omega\right)\right]\\ \nonumber
                              &\times&K\left(\omega,\omega^{'}\right)\phi\left(\omega+\omega^{'}\right)],
\end{eqnarray}
and
\begin{eqnarray}
\label{r2}
Z\left(\omega\right)&=&
                                  1+\frac{i\pi}{\omega\beta}\sum_{m=-M}^{M}
                                  \frac{\lambda\left(\omega-i\omega_{m}\right)\omega_{m}}{\sqrt{\omega_m^2Z^{2}_{m}+\phi^{2}_{m}}}Z_{m}\\ \nonumber
                              &+&\frac{i\pi}{\omega}\int_{0}^{+\infty}d\omega^{'}\alpha^{2}F\left(\omega^{'}\right)
                                  [\left[N\left(\omega^{'}\right)+f\left(\omega^{'}-\omega\right)\right]\\ \nonumber
                              &\times&K\left(\omega,-\omega^{'}\right)\left(\omega-\omega^{'}\right)Z\left(\omega-\omega^{'}\right)]\\ \nonumber
                              &+&\frac{i\pi}{\omega}\int_{0}^{+\infty}d\omega^{'}\alpha^{2}F\left(\omega^{'}\right)
                                  [\left[N\left(\omega^{'}\right)+f\left(\omega^{'}+\omega\right)\right]\\ \nonumber
                              &\times&K\left(\omega,\omega^{'}\right)\left(\omega+\omega^{'}\right)Z\left(\omega+\omega^{'}\right)], 
\end{eqnarray}
where:
\begin{equation}
\label{r3}
K\left(\omega,\omega^{'}\right)\equiv
\frac{1}{\sqrt{\left(\omega+\omega^{'}\right)^{2}Z^{2}\left(\omega+\omega^{'}\right)-\phi^{2}\left(\omega+\omega^{'}\right)}}.
\end{equation}

The Eliashberg set, Eqs. (\eq{r1} and \eq{r2}), gives the following solutions on the real axis: the order parameter function $\phi\left(\omega\right)$ and the wave function renormalization factor $Z\left(\omega\right)$. The order parameter is defined by the expression: $\Delta\left(\omega\right)\equiv \phi\left(\omega\right)/Z\left(\omega\right)$. On the other hand, the imaginary axis functions ($\phi_{m}\equiv\phi\left(i\omega_{m}\right)$ and $Z_{m}\equiv Z\left(i\omega_{m}\right)$) should be calculated by using the equations: 
\begin{equation}
\label{r4}
\phi_{n}=\frac{\pi}{\beta}\sum_{m=-M}^{M}
\frac{\lambda\left(i\omega_{n}-i\omega_{m}\right)-\mu^{*}\left(\omega_{m}\right)}
{\sqrt{\omega_m^2Z^{2}_{m}+\phi^{2}_{m}}}\phi_{m},
\end{equation}
and
\begin{equation}
\label{r5}
Z_{n}=1+\frac{1}{\omega_{n}}\frac{\pi}{\beta}\sum_{m=-M}^{M}
\frac{\lambda\left(i\omega_{n}-i\omega_{m}\right)}{\sqrt{\omega_m^2Z^{2}_{m}+\phi^{2}_{m}}}
\omega_{m}Z_{m},
\end{equation}
where the Matsubara energy is given by: $\omega_{m}\equiv \left(\pi / \beta\right)\left(2m-1\right)$ and $\beta\equiv\left(k_{B}T\right)^{-1}$; $k_{B}$ is the Boltzmann constant.

The pairing kernel for the electron-phonon interaction has the form:
\begin{equation}
\label{r6}
\lambda\left(z\right)\equiv 2\int_0^{\Omega_{\rm{max}}}d\Omega\frac{\Omega}{\Omega ^2-z^{2}}\alpha^{2}F\left(\Omega\right).
\end{equation}
The Eliashberg function for the molecular hydrogen ($\alpha^{2}F\left(\Omega\right)$) has been calculated by L. Zhang {\it et al.} \cite{Zhang}. The maximum phonon energy ($\Omega_{\rm{max}}$) is equal to $508.1$ meV.

The function $\mu^{*}\left(\omega_{m}\right)\equiv\mu^{*}\theta\left(\omega_{c}-|\omega_{m}|\right)$ describes the Coulomb repulsion between electrons;  $\mu^{*}$ denotes the Coulomb pseudopotential, $\theta$ is the Heaviside unit function and $\omega_{c}$ represents the energy cut-off ($\omega_{c}=3\Omega_{\rm{max}}$). For the molecular hydrogen we assume: $\mu^{*}\in\left<0.08,0.15\right>$.  

The symbols $N\left(\omega\right)$ and $f\left(\omega\right)$ denote the Bose and Fermi functions respectively.

The Eliashberg equations have been solved for $1601$ Matsubara frequencies ($M=800$) by using the numerical method presented in the paper \cite{Szczesniak1} and \cite{Szczesniak2}. In the considered case, the functions $\phi\left(\omega\right)$ and $Z\left(\omega\right)$ are stable for $T\geq 11.6$ K.

\section{Results}

%
\begin{figure}[t]%
\includegraphics*[scale=0.31]{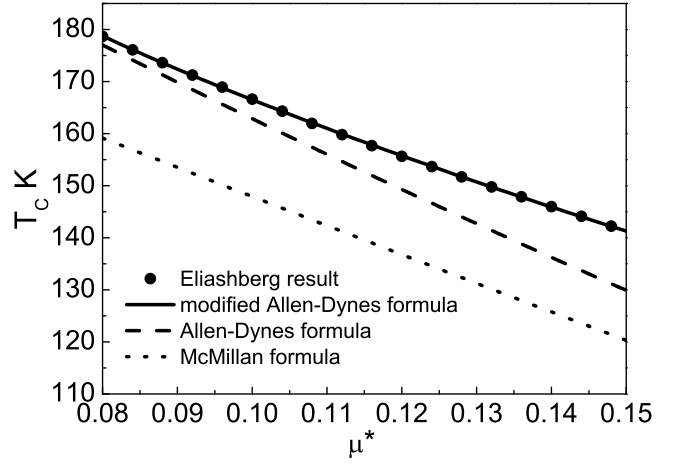}
\caption{The critical temperature as a function of the Coulomb pseudopotential.}
\label{f1} 
\end{figure}
%
\begin{figure}[t]%
\includegraphics*[scale=0.31]{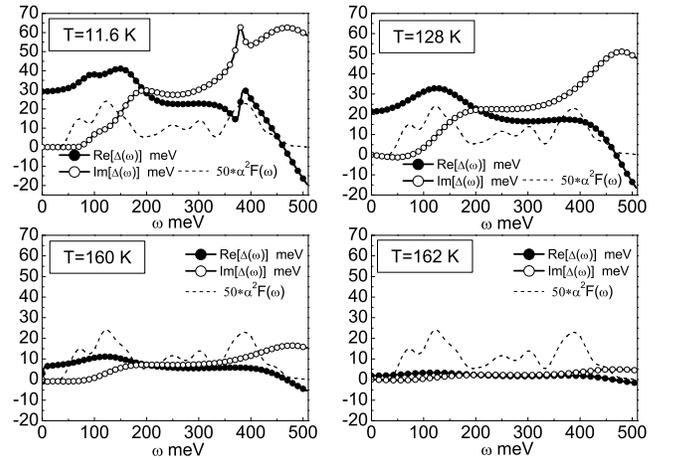}
\caption{
The real and imaginary part of the order parameter on the real axis for the selected temperatures. The rescaled Eliashberg function is also plotted.} 
\label{f2}
\end{figure}
%
\begin{figure}[t]%
\includegraphics*[scale=0.33]{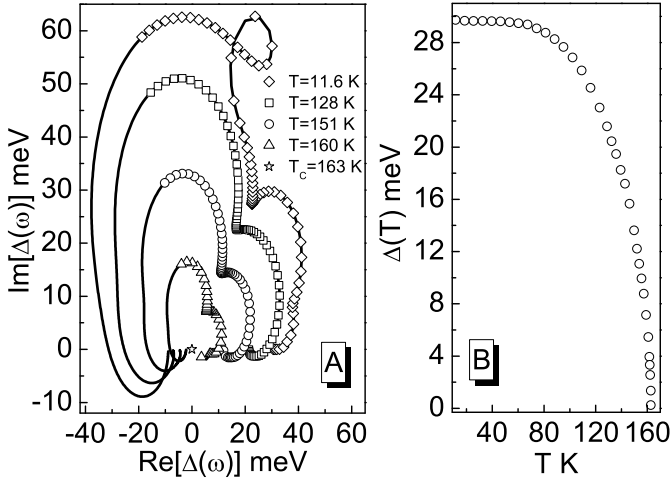}
\caption{(A) The order parameter on the complex plane for the selected values of the temperature. The lines with symbols represent the solutions for $\omega\in\left<0,\Omega_{\rm max}\right>$, whereas the regular lines correspond to the solutions for $\omega\in\left(\Omega_{\rm max}, \omega_{c}\right>$. (B) The temperature dependence of the order parameter. In the both cases we assume $\mu^{*}=0.1$.} 
\label{f3}
\end{figure}
%


In Fig.\fig{f1} we have presented the dependence of the critical temperature on the value of the Coulomb pseudopotential. The exact numerical solutions of the Eliashberg equations have been represented by the black circles. The dotted and dashed lines represent the calculation of the critical temperature by   using the McMillan formula and the more elaborated Allen-Dynes expression \cite{AllenDynes} respectively. It is easy to see, that McMillan formula lowers much $T_{C}$ in the whole range of the Coulomb pseudopotential's values, whereas the Allen-Dynes expression predicts correctly the critical temperature only for very low values of $\mu^{*}$. For this reason we have modified the classical Allen-Dynes expression in order to obtain the analytical formula, which reproduces the Eliashberg results exactly (the solid line in Fig.\fig{f1}). In particular, we have fitted the selected parameters in the Allen-Dynes expression ($\Lambda_{1}$ and $\Lambda_{2}$) with the help of $250$ numerical values of $T_{C}\left(\mu^{*}\right)$. The final result takes the form:    
\begin{equation}
\label{r7}
k_{B}T_{C}=f_{1}f_{2}\frac{\omega_{\rm ln}}{1.2}\exp\left[\frac{-1.04\left(1+\lambda\right)}{\lambda-\mu^{*}\left(1+0.62\lambda\right)}\right],
\end{equation}
where:
\begin{equation}
\label{r8}
f_{1}\equiv\left[1+\left(\frac{\lambda}{\Lambda_{1}}\right)^{\frac{3}{2}}\right]^{\frac{1}{3}}, \quad {\rm and} \quad
f_{2}\equiv 1+\frac{\left(\frac{\sqrt{\omega_{2}}}{\omega_{\rm{ln}}}-1\right)\lambda^{2}}{\lambda^{2}+\Lambda^{2}_{2}}.
\end{equation}
Additionally:
\begin{equation}
\label{r9}
\Lambda_{1}\equiv 3.64-12.92\mu^{*}, \quad {\rm and} \quad
\Lambda_{2}\equiv \frac{\sqrt{\omega_{2}}}{\omega_{\rm{ln}}}\left(1.39-59.74\mu^{*}\right).
\end{equation}
The parameters $\lambda$, $\sqrt{\omega_{2}}$ and $\omega_{{\rm ln}}$ are equal to $1.2$, $207.5$ meV and $141.9$ meV, respectively.


In Fig.\fig{f2} we have shown the order parameter on the real axis for the selected temperatures, and $\mu^{*}=0.1$; the Eliashberg function is also plotted. On the basis of the presented results one can state, that both the real and imaginary part of the function $\Delta\left(\omega\right)$ is plainly correlated with the shape of the electron-phonon interaction. This effect is especially clearly visible for the low values of the temperature. The full form of the order parameter on the complex plane has been presented in Fig.\fig{f3} (A). We have stated, that the $\Delta\left(\omega\right)$ values form the distorted spirals which shrink with the growth of the temperature. Basing on Fig.\fig{f3} (A) it is also possible to notice, that the effective electron-electron interaction is attractive (Re$\left[\Delta\left(\omega\right)\right] > 0$) in the range of the frequencies from zero to $\sim 0.9\Omega_{{\rm max}}$.     

\begin{figure}[t]%
\includegraphics*[scale=0.31]{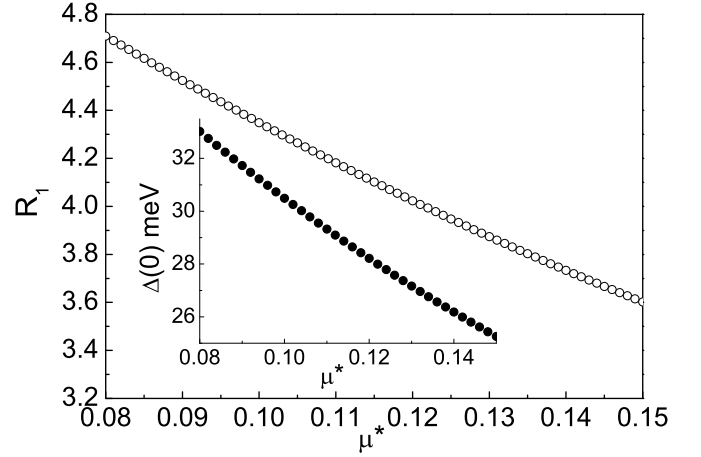}
\caption{The ratio $R_{1}$ as a function of the Coulomb pseudopotential. The inset shows the open form of $\Delta\left(0\right)$.}
\label{f4}
\end{figure}

Taking into consideration the equation: $\Delta\left(T\right)={\rm Re}\left[\Delta\left(\omega=\Delta\left(T\right)\right)\right]$ we have calculated the dependence of the order parameter on the temperature (see Fig.\fig{f3} (B)). Next, the value of the ratio $R_{1}\equiv\frac{2\Delta\left(0\right)}{k_{B}T_{C}}$ can be obtained, where $\Delta\left(0\right)$ denotes the value of the order parameter close to the zero temperature and $\Delta\left(0\right)\simeq\Delta\left(T=11.6 {\rm K}\right)$. In Fig.\fig{f4} we have presented the possible values of $R_{1}$ for $\mu^{*}\in\left<0.08,0.15\right>$. It is easy to see, that in the whole range of the considered values of the Coulomb pseudopotential, the ratio $R_{1}$ differs essentially from the BCS prediction; $\left[R_{1}\right]_{\rm BCS}=3.53$ \cite{BCS}. Additionally, in the inset in Fig.\fig{f4} we have shown the dependence of $\Delta\left(0\right)$ on $\mu^{*}$.
      
%
%
\begin{figure}[t]%
\includegraphics*[scale=0.31]{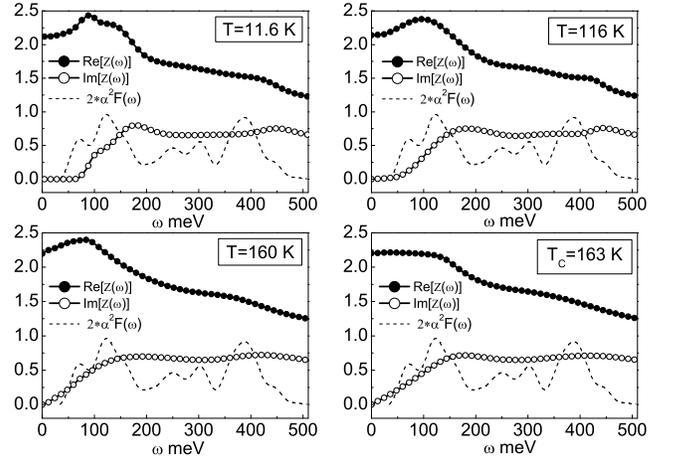}
\caption{The real and imaginary part of the wave function renormalization factor on the real axis for the selected temperatures. The rescaled Eliashberg function is also plotted.}
\label{f5}
\end{figure}
%
\begin{figure}[t]%
\includegraphics*[scale=0.31]{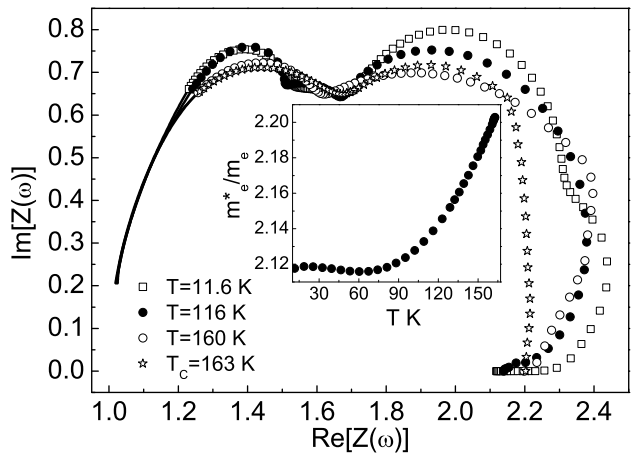}
\caption{The wave function renormalization factor on the complex plane ($\mu^{*}=0.1$). The lines with symbols represent the solutions for $\omega\in\left<0,\Omega_{\rm max}\right>$, whereas the regular lines correspond to the solutions for $\omega\in\left(\Omega_{\rm max}, \omega_{c}\right>$. The inset shows the ratio $m^{*}_{e}/m_{e}$ as a function of the temperature.}
\label{f6}
\end{figure}
%

The second solution of the Eliashberg equations on the real axis has been plotted in Fig.\fig{f5} ($\mu^{*}=0.1$). The obtained results prove, that the function $Z\left(\omega\right)$ also clearly senses the structure of the electron-phonon interaction. The shape of the wave function renormalization factor on the complex plane has been presented in Fig.\fig{f6}. We see, that in contrast to the order parameter, the function $Z\left(\omega\right)$ weakly depends on the temperature.  

In the framework of the Eliashberg formalism, on the basis of the wave function renormalization factor, the temperature dependence of the electron effective mass ($m^{*}_{e}$) can be calculated. In particular:  $m^{*}_{e}/m_{e}={\rm Re}\left[Z\left(0\right)\right]$, where the symbol $m_{e}$ denotes the bare electron mass. The values of the ratio $m^{*}_{e}/m_{e}$ have been presented in the Fig's.\fig{f6} inset. According to the presented data, it is easy to spot, that $m^{*}_{e}$ is high in the full range of the considered temperatures and $\left[m^{*}_{e}/m_{e}\right]_{\rm max}=2.2$ for $T=T_{C}$. We notice, that at the critical temperature, the electron effective mass is independent of $\mu^{*}$.

\section{Summary}

We have calculated the selected thermodynamic properties of the superconducting state in the molecular hydrogen ($p=428$ GPa). The numerical analysis has been conducted in the framework of the one-band Eliashberg formalism for the wide range of the Coulomb pseudopotential's values: $\mu^{*}\in\left<0.08,0.15\right>$. We have proved, that the critical temperature is high even for large values of $\mu^{*}$  ($\left[T_{C}\right]_{\rm min}=141$ K). Next, it has been shown, that the superconducting phase is characterized by high value of the dimensionless ratio $R_{1}$, which differs from the BCS result; $R_{1}\in\left<4.71, 3.60\right>$. Finally, we have observed, that the electron-phonon interaction strongly enhances the effective electron mass for the temperatures from zero to $T_{C}$. In particular, the maximum of $m^{*}_{e}$ is equal to $2.2m_{e}$ for $T= T_{C}$.

In the future we will analyze the superconducting state in the molecular hydrogen in the framework of the three-gap Eliashberg model. It will be done in order to discuss the effect of the multi-band anisotropy \cite{Cudazzo}.
  
\begin{acknowledgments}

The authors wish to thank K. Dzili{\'n}ski for providing excellent working conditions and the financial support; A.P. Durajski and 
D. Szcz{\c{e}}{\'s}niak for the productive scientific discussion that improved the quality of the presented work. All numerical calculations were based on the Eliashberg function sent to us by: {\bf L. Zhang}, Y. Niu, Q. Li, T. Cui, Y. Wang, {\bf Y. Ma}, Z. He and G. Zou for whom we are also very thankful.  

\end{acknowledgments}

%
\end{document}